# Exoplanets versus brown dwarfs : the CoRoT view and the future[1]

*Jean Schneider*
*LUTh – Observatoire de Paris*

CoRoT has detected by transit several tens of objects (Moutou & Deleuil 2015) whose radii run from 1.67 Earth radius (CoRoT 7b, Leger et al. 2009) to 1.5 Jupiter radius (CoRoT-1 b, Barge et al. 2007) [2]. Their mass run from less than 5.7 Earth mass (CoRoT-24 b, Alonso et al. 2014) to 63 Jupiter mass (CoRoT-15 b, Bouchy et al. 2011). Their mass-radius diagram is represented in Figure 1 below. One could be tempted to think that more massive the object is, the larger it is in size and that there is some limit in mass and/or radius beyond which objects are not planets but very low mass stars below the 80 Jupiter mass limit to trigger nuclear fusion (namely « brown dwarfs » ). CoRoT findings contribute to the planet versus brown dwarf debate since the Figure 1 shows that there is no clear mass-radius relation.

2 Data as of October 15 2015

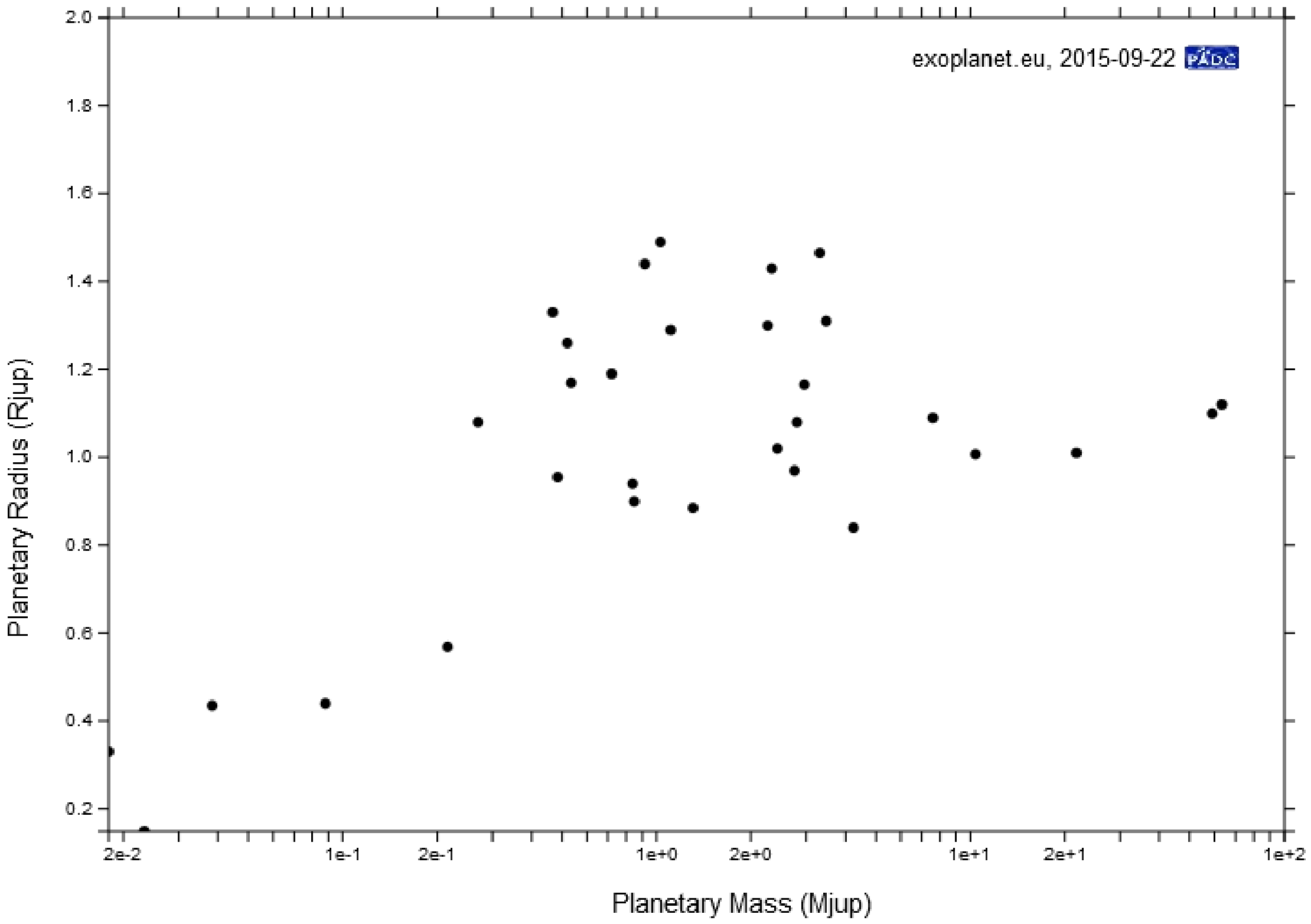

**Figure 1.** *Mass radius relation for CoRoT objects (15 Oct 2015) from exoplanet.eu.*

One is thus facing two problems : terminology (what is a planet ? what is a brown dwarf?) and classification (how to decide if a given object is a planet or a brown dwarf according to a given definition ?). Let us discuss these two aspects and the CoRoT contribution.

**What is a planet ?**

The debate, open by several authors (see for example Baraffe et al. 2010, Schneider et al. 2011, Hatzes & Rauer 2015), is still ongoing and will not be closed by the present contribution. Names are arbitrary conventions, but the naturel trend is to make them designate sufficiently elaborated concepts. Derived from the solar system analogy, exoplanets (in short planets) designate small bodies orbiting around stars and formed by condensation in a circumstellar dust disk. A first question is of course how small has the body to be for being a planet. The problem here is that there exist small bodies orbiting stars which are probably not formed like planets, namely brown dwarfs forming, like stars, by collapse of a (possibly dusty) gas cloud.

**From the heaven of concepts to the hell of observations**

So we have a clear conceptual discrimination between planets and brown dwarfs (keeping in mind that it is a convention). But it is based on a criterion involving an inobsrvable concept, namely its formation scenario, because we do not have the formation movie at hand. We can only rely on actual observables. Standard basic bulk observables are the object mass, radius, temperature. An ideal situation would be that at least for one of these observables there exist two domains $D_{planet}$ and $D_{brown\ dwarf}$ of values which do not intersect. It is unfortunately not the case since there are planets smaller or larger, heavier or lighter, cooler or hotter than objects we believe to be brown dwarfs, Even worse, there are a few pulsar companions with masses lower than 30 Jupiter mass.  They are probably the relict of stellar companions eroded by the pulsar strong wind (Ray & Loeb 2015). One can argue that as such they are not planets nor brown dwarfs, their formation process being very different. But one cannot exclude that such erosion mechanism happend also for low mass compnions of main sequence stars with strong winds (see *e.g.* Sanz-Forceda et al. 2010). The choice made by the Extrasolar Encyclopaedia at exoplanet.eu, based on Hatzes & Rauer (2015), is to take all objects below 60 Jupiter mass.

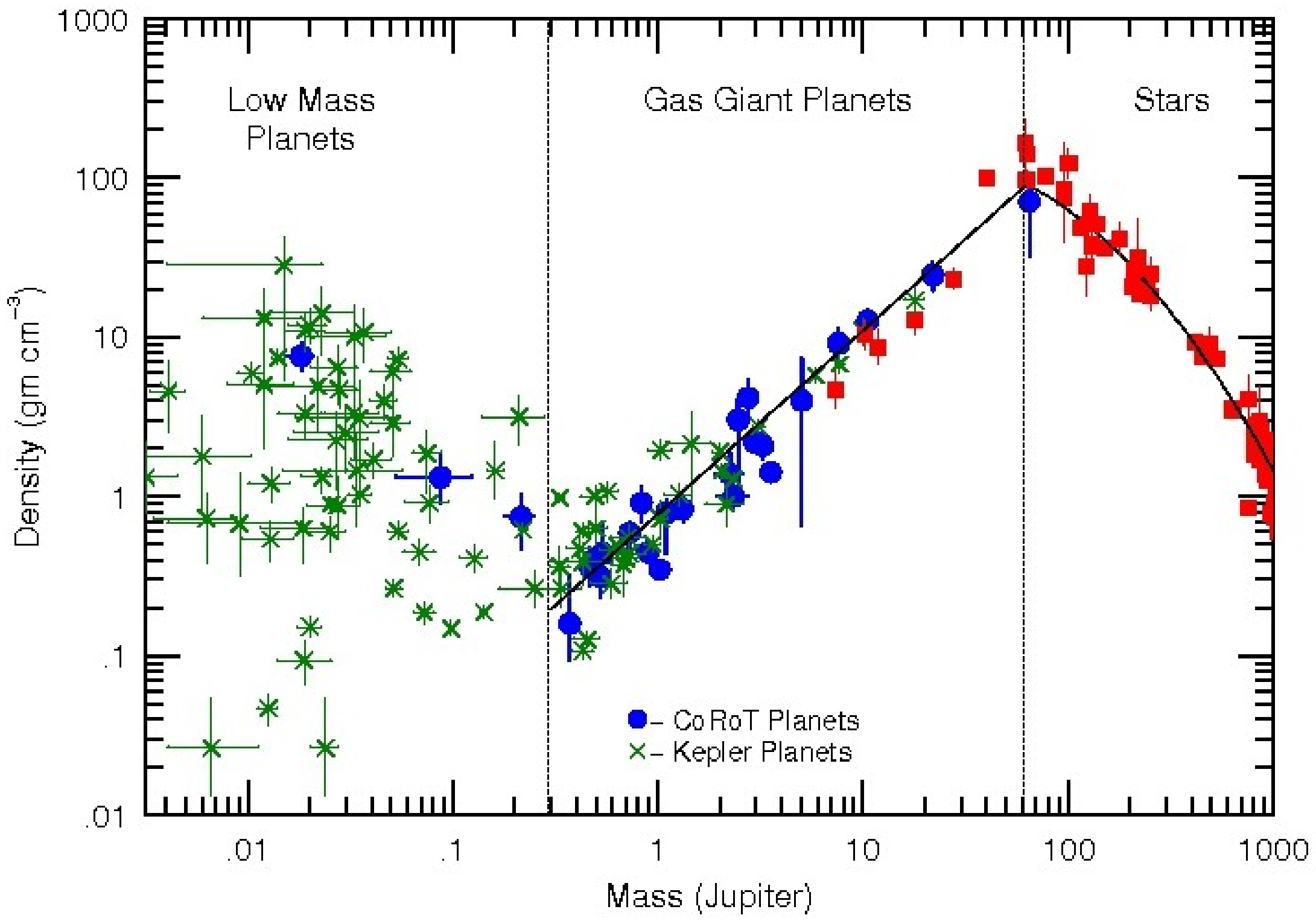

**Figure 2** *Empirical mass-density relation (Hatzes & Rauer 2015)*

The Hatzes & Rauer argument is that the mass-radius and the mass-density relation presents no particular feature in the giant planet régime (i.e. more massive than Saturn) and that there is a change in the slope of distribution at 60 Jupiter mass (Figure 2). But unfortunately their statistics in the 30-60 Jupiter mass region is poor (the so called brown dwarf desert) since they rely only on transiting planets and the authors do no consider the mass histogramme in this region. Earlier data suggested a dip around 40 Jupiter mass (Sahlman et al. 2011, Udry 2010 – Figures 3 and 4) in the mass histogramme. More statistics will come in the near future including radial velocity data from rge ground and astrometric data from Gaia to see if a feature around 40 Jupiter mass in the mass-radius diagramme exists or not.

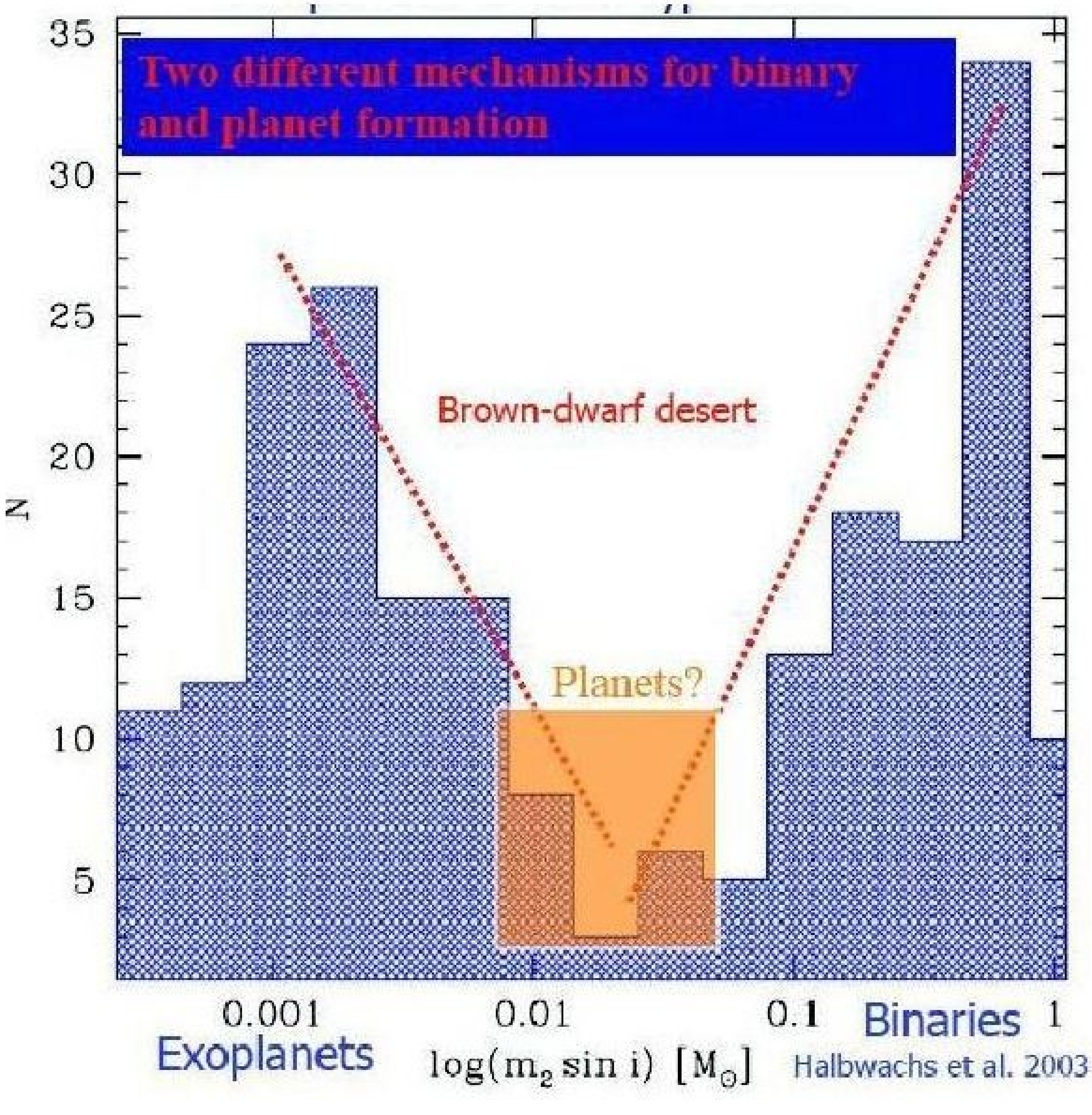

**Figure 3** *Mass histogramme of low mass objects (Udry et al. 2010)*

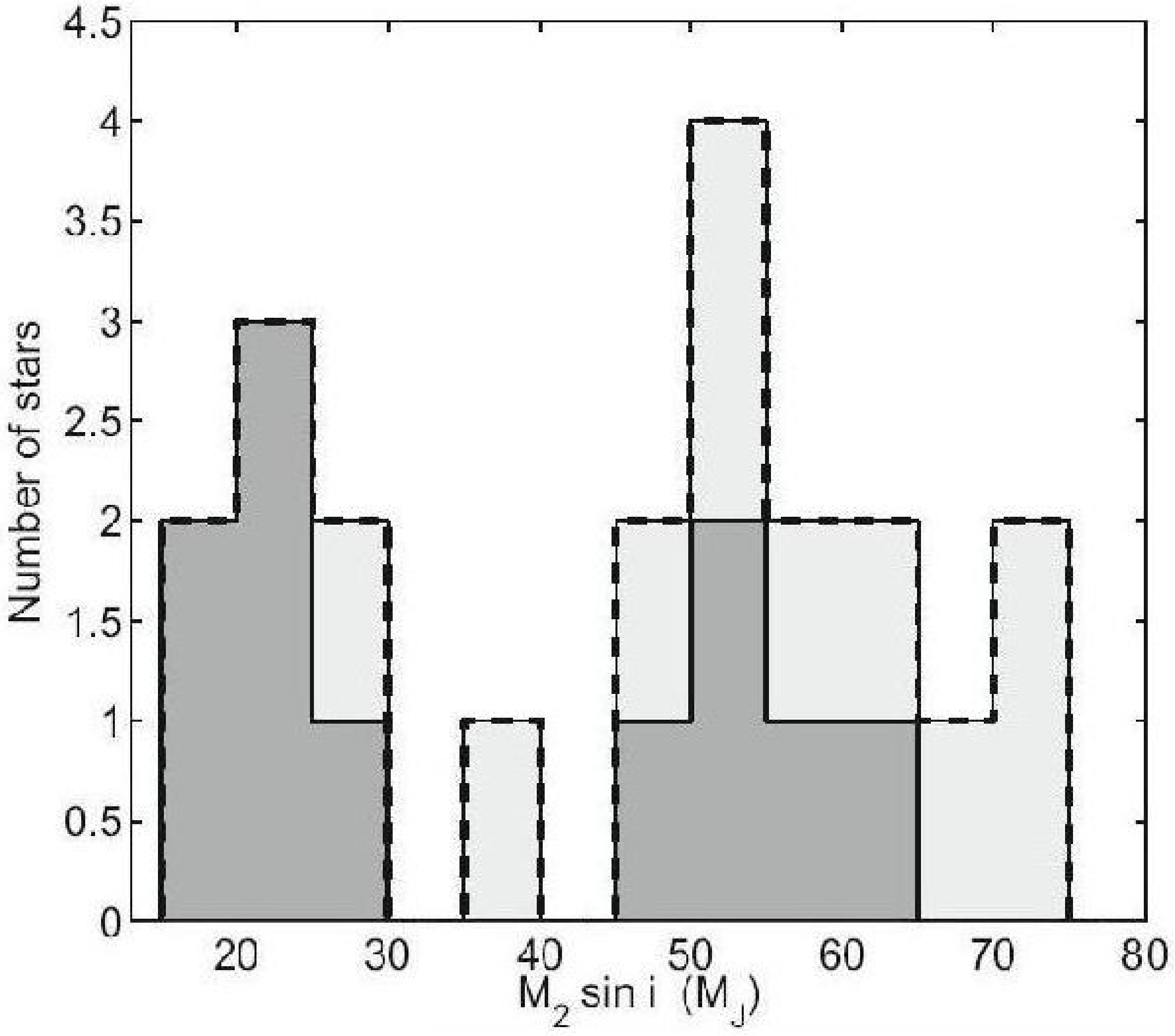

**Figure 4** *Low mass objects histogramme in the 20 – 75 Jupiter mass region (Sahlman et al. 2011)*

A future improvement to separate the planet and brown dwarf populations will come from advanced observables, like the spectral type and species composition. They will help to constrain the formation mechanism of the object (accretion in a dust disk or collapse of a gas cloud).

At least one conclusion is clear, the former mass limit of exoplanets at 13 Jupiter mass, correspinding to the triggering of nuclear burning of deuterium, is not relevant since an object can be formed by dust accretion and acquire a final mass larger than 13 Jupiter.

There is a second, more factual, problem : the value of observables can be very uncertain. This is especially the case for objects detected by imaging where the mass cannot be infered from radial velocity measurements but only from spectra and models. A typical example is the object 2M1207 (Chauvin et al. 2005) with a Jupiter mass of 4 ± 1 Jupiter mass can be derived from its spectra. Indeed, in these cases the star-planet separation is so wide that the semi-amplitude $K=\sqrt{(GM_{star}/a_{planet})}$ of the stellar radial velocity variation induced by the planet motion is too low to be measurable. Even more : when the mass determination is as precise as a percent (in case of radial velocity or astrometric measurements), one faces the absurd situation of a sharp mass limit. For example, what to do with objects like CoRoT-15 b with M = 63.3 ± 4 $M_{Jup}$ ?

A last problem, which we do not address here because the concerned population is generally supposed to be small, is the « intersteller wanderers », i.e. planets ejected by dynamical interaction from a well formed planetary system.

## Conclusion

Assuming that the definition of a planet and a brown dwarf is adopted according to their formation mechanism, to separate the two populations is not an easy task. Any catalogue contains necessarily a mixture of both populations. Since catalogues are useful not only to list characteristics of objects but also to make statistics on these characteristics, I recommend to take low constrains (for our case a mass limit as high as 60 Jupiter mass) on the properties used to define a sample, in order not to miss interesting objects. Modern softwares used to read electronic catalogues allow to eliminate easily objects from a catalogue which do not fullfil the criteria of definition of each user, which is free to impose his own criteria.